\documentclass[twocolumn,superscriptaddress,showpacs,letterpaper]{revtex4-1}%
\makeatletter

\newcommand{\Rmnum}[1]{\expandafter\@slowromancap\romannumeral #1@}
\makeatother
\usepackage{graphicx}
\usepackage{sidecap}
\usepackage{latexsym}
\usepackage{amssymb}
\usepackage{amsfonts}
\usepackage{soul}
\usepackage{color}
\usepackage{amsmath}
\usepackage{epstopdf}
\usepackage{mathrsfs}
\usepackage{bm}
\usepackage{float}

\begin{document}

\title{Holistic numerical simulation of a quenching process on a \\ real-size multifilamentary superconducting coil}

\author{Cun Xue}
\email{xuecun@nwpu.edu.cn}
\affiliation{School of Mechanics, Civil Engineering and Architecture, Northwestern Polytechnical University, Xi'an 710072, China}
\author{Han-Xi Ren}
\affiliation{School of Aeronautics, Northwestern Polytechnical University, Xi'an 710072, China}
\author{Peng Jia}
\affiliation{School of Mechanics, Civil Engineering and Architecture, Northwestern Polytechnical University, Xi'an 710072, China}
\author{Qing-Yu Wang}
\affiliation{School of Aeronautics, Northwestern Polytechnical University, Xi'an 710072, China}
\author{Wei Liu}
\affiliation{Western Superconducting Technologies Co., Ltd., and Xi'an Superconducting Magnet Technology Co., Ltd, Xi'an 710014, China}
\author{Xian-Jin Ou}
\affiliation{Institute of Modern Physics, Chinese Academy of Sciences, Lanzhou 730000, China}
\author{Liang-Ting Sun}
\affiliation{Institute of Modern Physics, Chinese Academy of Sciences, Lanzhou 730000, China}
\affiliation{School of Nuclear Science and Technology, University of Chinese Academy of Sciences, Beijing 100049, China}
\author{Alejandro V Silhanek}
\affiliation{Experimental Physics of Nanostructured Materials, Department of Physics, Universit\'{e} de Li\`{e}ge, B-4000 Sart Tilman, Belgium}

\date{\today}

\begin{abstract}

Superconductors play a crucial role in the advancement of high-field electromagnets. Unfortunately, their performance can be compromised by thermomagnetic instabilities, wherein the interplay of rapid magnetic and slow heat diffusion can result in catastrophic flux jumps eventually leading to irreversible damage. This issue has long plagued high-$J_c$ Nb$_3$Sn wires at the core of high-field magnets. In this study, we introduce a groundbreaking large-scale GPU-optimized algorithm aimed at tackling the complex intertwined effects of electromagnetism, heating, and strain acting concomitantly during the quenching process of superconducting coils. We validate our model by conducting comparisons with magnetization measurements obtained from short multifilamentary Nb$_3$Sn wires and further experimental tests conducted on solenoid coils while subject to ramping transport currents. Furthermore, leveraging our developed numerical algorithm, we unveil the dynamic propagation mechanisms underlying thermomagnetic instabilities (including flux jumps and quenches) within the coils. Remarkably, our findings reveal that the velocity field of flux jumps and quenches within the coil is correlated with the amount of Joule heating experienced by each wire over a specific time interval, rather than solely being dependent on instantaneous Joule heating or maximum temperature. These insights have the potential to pave the way for optimizing the design of next-generation superconducting magnets, thereby directly influencing a wide array of technologically relevant and multidisciplinary applications.

\end{abstract}

\maketitle

Due to high current carrying capability with loss-less characteristic, superconductors are essential components for the development of high-field electromagnets. However, their performance can be threatened by thermomagnetic instabilities, a phenomenon in which the interplay between swift flux motion and slow heat diffusion give rise to sudden flux bursts which limit the lifetime of the coil. Indeed, frequent magnetic flux jumps had been identified as a long standing issue \cite{FJ-MIT1,FJ-Ohio1,FJ-Ohio2,FJ-LBNL1,FJ-Italy1,FJ-CERN1,FJ-Switzerland1,FJ-Switzerland2,FJ-Fermilab1} at the source of serious problems in high-$J_c$ Nb$_3$Sn wires/strands used in 10-16 T magnets \cite{IMP} and over-20 T hybrid magnets \cite{wangqiuliang}. Previous reports have shown that flux jumps may cause premature quenches at low-fields and currents well below the designed operating regime \cite{BNL-Ghosh1,BNL-Ghosh2,Fermilab-Zolbin1,Fermilab-Zolbin2}. In this case, a rather lengthy, helium-intensive, and expensive process of magnet training is needed in order to achieve the targeted maximum field and ramp rate \cite{long-training}. Additionally, the stochastic behavior of magnetic flux jumps significantly affects the field stability in the magnet bore and makes accurate field-correction protocols particularly challenging \cite{field-correction1,field-correction2}. Furthermore, prevention measures via quench detection systems based on voltage spikes seem to be prone to errors \cite{field-correction1,QDS-noise1}.

Soon after magnetic flux jumps were first observed and investigated in 1960s \cite{FJ-discovery}, the underlying physical mechanism were revealed \cite{FJ-physics1,FJ-physics2,FJ-physics3,FJ-physics4,FJ-physics5,FJ-physics6,FJ-physics7,FJ-physics8} along with the relevant physical parameters (temperature \cite{FJ-temperature}, ramping rate \cite{FJ-rampingrate}, sample size \cite{FJ-samplesize}, border defects \cite{FJ-defects}) ruling the nucleation and growth of thermomagnetic instabilities. For composite superconducting wires/strands, early criteria for triggering magnetization flux jumps were proposed by Swartz \& Bean \cite{FJ-wire-Swartz} and Wilson \cite{FJ-wire-Wilson}. Subsequently, a series of studies were carried out to describe the characteristics of low-field flux jumps in order to develop a new generation of Nb$_3$Sn high-field magnets \cite{FJ-wire-theory1,FJ-wire-theory2,FJ-wire-theory3,FJ-wire-theory4,FJ-wire-theory5}. It was found that reducing the effective filament size and improving the residual resistivity ratio (RRR) are of paramount importance for suppressing flux jumps \cite{FJ-wire-theory2, FJ-RRR-size1,FJ-RRR-size2}. More recently, Xu et al. investigated the influence of heat treatment temperature and Ti-doping on flux jumps and demonstrated that introducing high specific heat substances can improve the stability of Nb$_3$Sn wires \cite{FJ-Xu1,FJ-Xu2}. However, thus far, all efforts have been directed to single superconducting wires and most criteria have been established by general electromagnetic analysis in limited cases of adiabatic or isothermal assumptions. Unfortunately, the theoretical development for a single wire falls short to describe complex coils due to distinct characteristics of the latter. Namely, (i) Different wires in the coil are generally exposed to different ramping magnetic fields (see Fig. S1 in Section I of Supplementary Information). (ii) Wires in a coil are not isolated but rather represent complex correlated systems. (iii) The stability of each wire strongly depends on its time-dependent electromagnetic penetration as well as the thermal shock from neighboring wires during the occurrence of localized flux jumps. Consequently, a physically-grounded onset criterion for triggering flux jumps which is accurate for an isolated single wire, may no longer be applicable to a cutire coil. To date, there are not powerful-enough tools based on numerical algorithms or available commercial software able to deal with correlated systems such as those of technologically relevant coils typically involving thousands of multifilamentary wires. In this case, an optimal design from filament to global structure is still considered a daunting, if not impossible, task.

As a matter of fact, numerical simulations of the thermomagnetic instabilities leading to partial flux jumps or complete quenching of a full-sized coil represent a formidably complex quest due to several reasons. Firstly, the relation between electric field $E$ and current density $J$ exhibits a very strong nonlinear dependence caused by the intricate flux dynamics involving enormous amount of nanoscale superconducting vortices. Secondly, the superconducting coils require a multiphysics approach including an interplay of heat diffusion, electromagnetic response and mechanical strain. Thirdly, unlike single phase superconducting samples (either in bulk or film form), the multiscale structures of magnets containing micro filaments, millimetric wires and metric coils, cannot be simulated through homogenization methods. Additionally, the sublements in each wire exhibit uncoupled electromagnetic responses for external magnetic field and coupled for transport current. Last, but not least, the thermal conductivity of cooper is 3-4 orders of magnitude larger than that of the epoxy. Since the dynamics of thermomagnetic instabilities relies on accurate temperature field calculations, it is impossible to obtain a satisfactory result for a composite coil simply by homogenization method with equivalent thermal parameters.

In this work, we develop an unprecedented large-scale GPU-advanced algorithm to address the aforementioned intractable problems of superconducting coils. We validate the model by comparing it with magnetization measurements of short multifilamentary Nb$_3$Sn wires and experimental tests performed on solenoid coils under a ramping transport current. Moreover, utilizing the developed numerical algorithm, we unveil the dynamic propagating processes of the thermomagnetic instabilities (flux jumps and quenches) in the coils. Surprisingly, we demonstrate that the velocity field of flux jumps and quenches in the coil result from the quantity of Joule heating in each wire within a certain time range rather than the instantaneous Joule heating and the maximum temperature. These results may provide the necessary breakthrough to optimize the design of next-generation superconducting magnets, with direct impact on technologically relevant and multidisciplinary applications.

\begin{figure} [t]
\centering \includegraphics[width=1.0\linewidth]{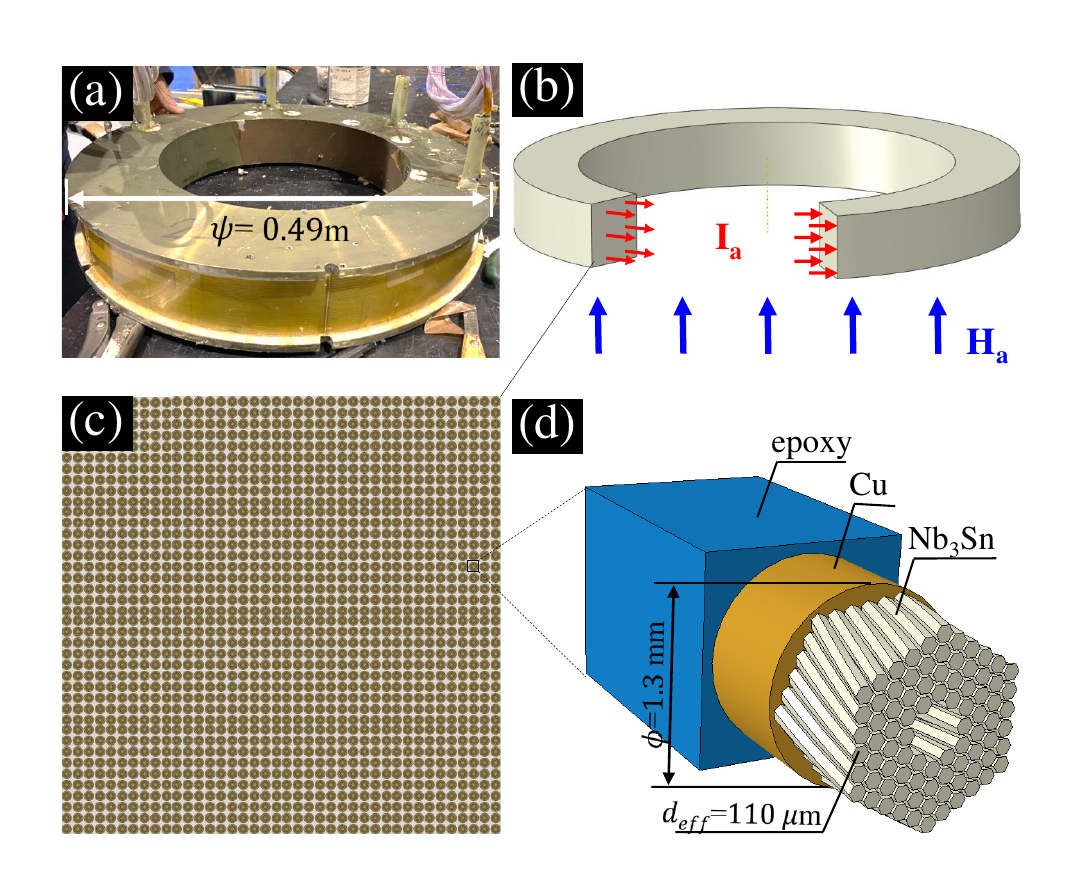}
\caption{(a) Solenoid superconducting magnet fabricated to benchmark against the numerical calculations. The solenoid consists of 1558 densely wound turns of high-$J_C$ Nb$_{3}$Sn wire with 84 subelements fabricated by internal-tin process (see details in section II of Supplementary Information). (b) Schematic of a solenoid coil exposed to a ramping transport current $I_a$ and a ramping external magnetic field $H_a$. (c-d) Cross-section of the solenoid coil with a zoom on the composite multifilamentary Nb$_{3}$Sn wire.}
\label{Fig:1}
\end{figure}

\begin{figure*} [ht]
\centering
\includegraphics*[width=0.8\linewidth,angle=0]{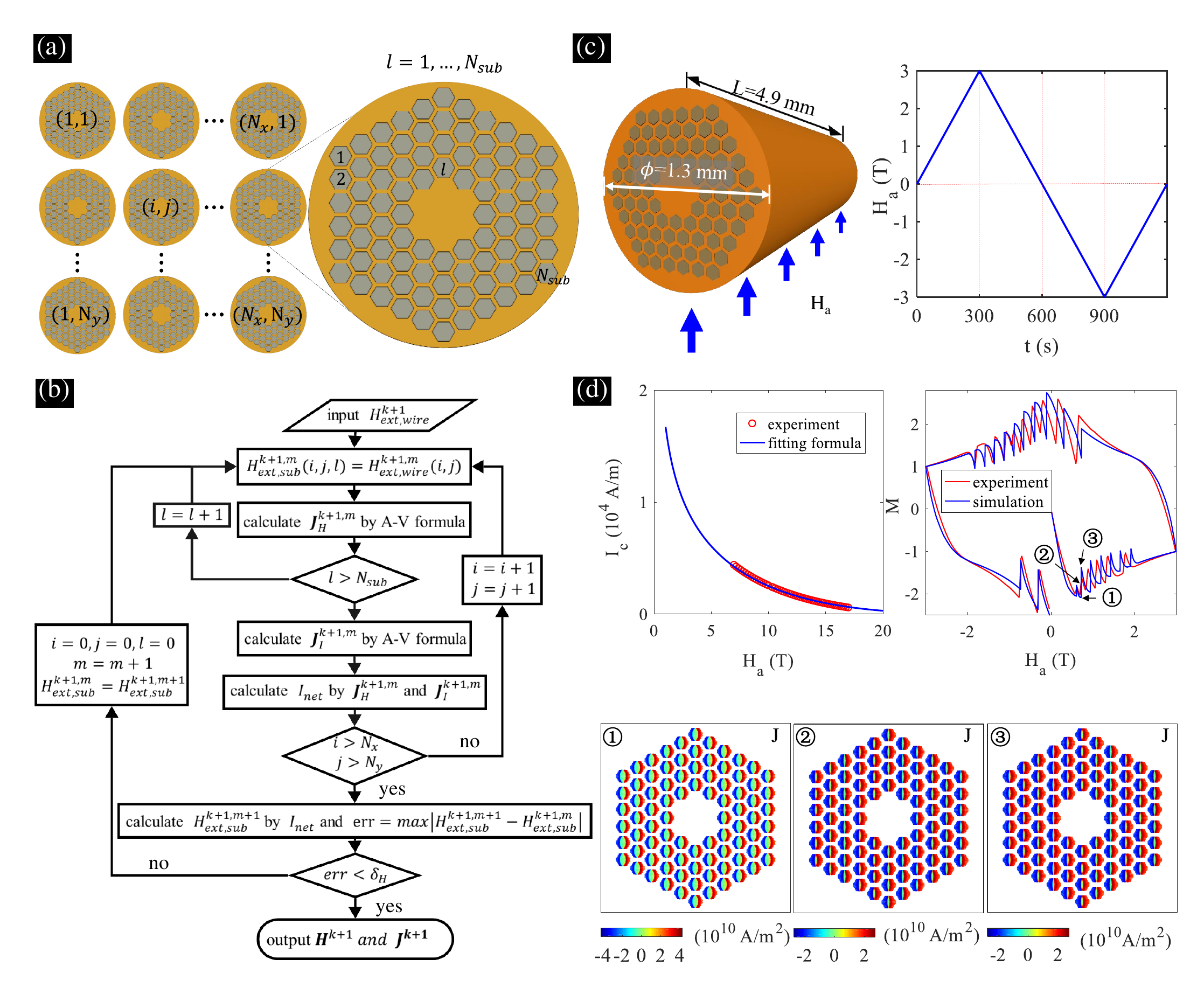}
\caption{ (a) Modeled system consisting of a superconducting coil with $N_x \times N_y$ turns. Each wire contains $N_{sub}$ subelements. (b) Flow chart for key subroutine in the numerical algorithm performed on graphics processing unit (GPU) to simulate the nucleation, growth and damping of thermomagnetic instabilities (flux jumps and quenches) in the superconducting coils. The flow chart for the main program can be seen in section IV of the Supplementary Information. (c) Schematic of the short segment of Nb$_{3}$Sn wire used to collect experimental measurements and exposed to an applied magnetic field $H_a(t)$. (d) (upper left) Variation of the critical current of the Nb$_{3}$Sn wire with magnetic field obtained by experimental measurements (red datapoints) and by fitting as described in the main text (blue curve). (upper right) Experimental and simulated magnetization of the Nb$_{3}$Sn short segment wire exposed to a transverse magnetic field loop with sweeping rate of 0.01 T/s at 4.2 K. The lower panels represents the simulated current density distributions during a flux jump.}
\label{Fig:2}
\end{figure*}

\begin{figure*} [ht]
\centering
\includegraphics*[width=0.8\linewidth,angle=0]{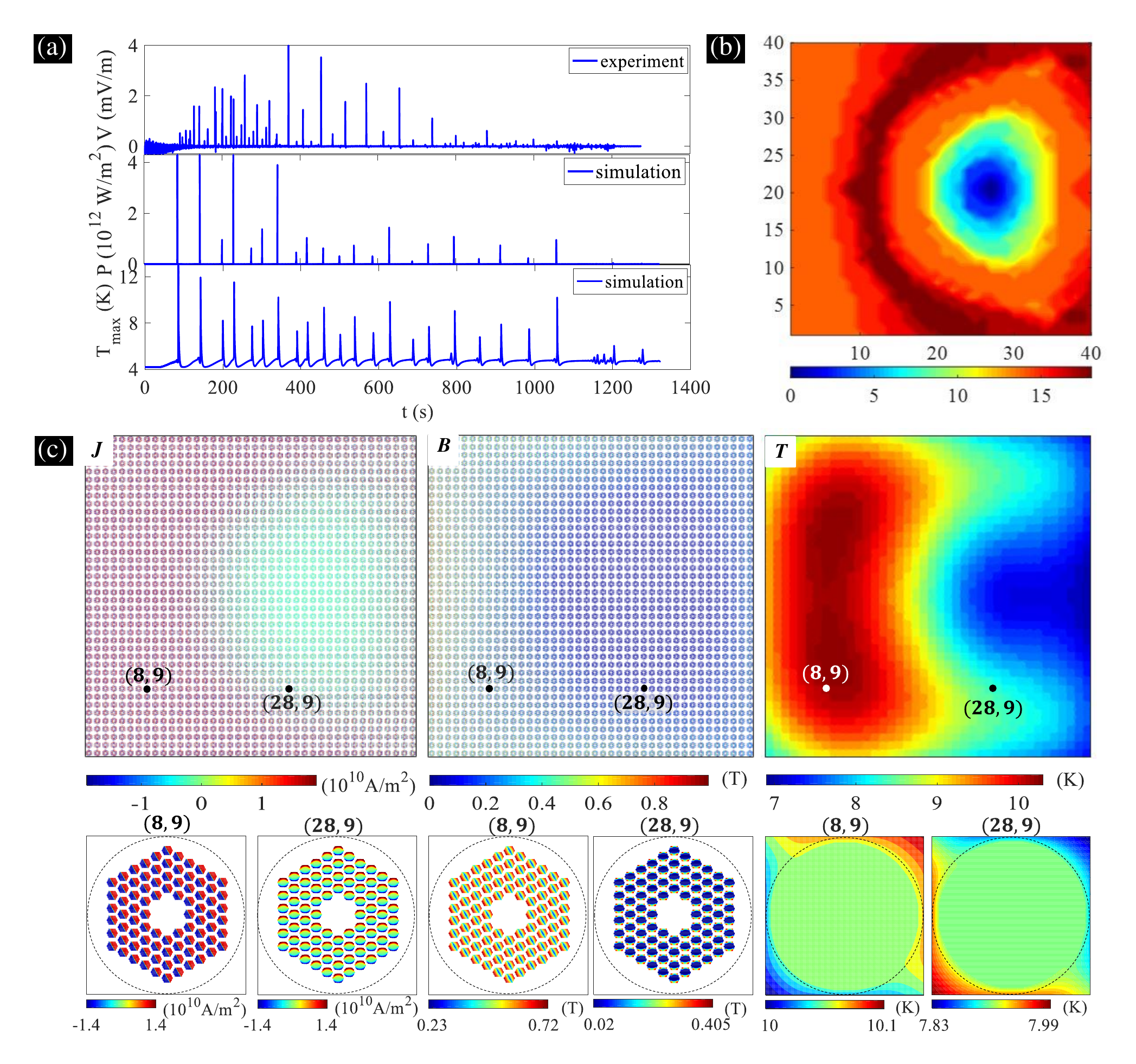}
\caption{(a) Voltage signal of a superconducting coil observed during an experimental test with ramping transport current of 0.5 A/s (upper panel). Simulated time-evolution of Joule heating power density and maximum temperature in the coil (lower two panels). (b) Contour-plot indicating the number of flux jumps for each wire during the ramping process. (c) The simulated current density, magnetic field and temperature during the second flux jump. The lower panels show detailed views for two wires with coordinates (8, 9) and (28, 9).}
\label{Fig:3}
\end{figure*}

\large
\vspace{4ex}
\noindent
\textbf{Methods}

\normalsize
\noindent
\textbf{Experiments}

In order to benchmark the numerical calculation against a real superconducting coil, we fabricated a solenoid consisting of 1558 ($38 \times 41$) turns of internal-tin (IT) Nb$_3$Sn wire, as shown in Fig. 1(a). The diameter of the bare Nb$_3$Sn wire is 1.3 mm. Each wire has 84 subelements and the averaged size of subelements is about 110 $\mu$m (Fig. 1 d). Each subelement consists of many filaments which are not drawn in the figure since they coalesce into a single mass. Indeed, in each of the internal-tin (IT) Nb$_3$Sn wire, the filaments coalesce to a continuous superconducting region within each subelement during a reactive heat treatment and thus the effective filament size $d_{eff}$ equals to the size of the entire subelement (filament is synonymous with subelement in this work). The ratio of copper (Cu) to Nb$_3$Sn is about $1.05$. Before cooling down to 4.2 K for experimental test, a pre-stress is induced to the solenoid coil by a preload layer by a thin aluminum strip. The solenoid magnet is then immersed in liquid helium inside a vacuum insulated Dewar permitting to keep the bath temperature at 4.2 K during the experimental test. Subsequently, the solenoid magnet is continuously fed with a transport current of increasing rate  0.5 A/s. The solenoid is only exposed to self-fields, without an external magnetic field. The maximum ramping rate of self-field in the coil during the test is about 7 mT/s.

\vspace{4ex}
\noindent
\textbf{Numerical Algorithm}

In order to explore the time-evolutions of thermomagnetic instabilities inside the superconducting coil, we develop a parallel numerical algorithm and run it on GPU. As shown in Fig. 1(b), we consider a solenoid coil wound by a multifilamentary superconducting wire, which is exposed to a ramping transport current $I_a$ and a ramping external magnetic field $H_a$. Due to the rotational symmetry of the solenoid coil, as shown in Fig. 1(b), it is sufficient to model a cross-section, as shown in Fig. 1(c). Without loss of generality, in the numerical model we have chosen to arrange the superconducting wires in a square array rather than the triangular array used in the real coil. Although we were unable to obtain strictly exact solutions for twisting superconducting wires with our numerical algorithm on the basis of the 2D model, our numerical model can capture the main twisting characteristics and provide very good approximate results with quite low error. Detail discussions of the twisting effect can be seen in Section IV of Supplementary Information.

Due to the fact that exists a resistive barrier between contiguous twisted subelements, when a non-current carrying Nb$_3$Sn wire is exposed to an external magnetic field, as shown in Fig. S8(b), negative and positive induced current circulate within each subelement whereas the net current of each subelement is zero (so-called uncoupled subelements). However, for a wire with transport current, the current density distributions in the entire cross-section of the wire is affected by the electromagnetic coupling between subelements, and in this case the transport current is first distributed in the outer subelements (completely coupled subelements). If a coil is fed with a ramping transport current, each wire undergoes a ramping transport current and a concomitant ramping external magnetic field generated by the other wires and coils in its vicinity. The fact that subelements of each wire are simultaneously uncoupled under external magnetic fields whereas they become completely coupled under applied transport current, represents a non-trivial problem to implement in the numerical simulations. Not less complex is the implementation of cross-talk of stray fields among nearby subelements.

Figure 2(a) graphically summarizes the numerical algorithm implemented in the present work. It consists of $N_x \times N_y$ turns in which each wire is labelled with a pair of coordinates ($i, j$) with $i=1 \ldots N_x$ and $j=1 \ldots N_y$. Each wire has $N_{sub}$ subelements. Both the turns of the coil and the number of subelements are parameters that can be adjusted in the numerical simulations. Fig. 2(b) shows the flow chart for the key subroutine of the numerical algorithm. In order to update the electromagnetic responses of a coil from the time step $k$ to the next time step $k+1$, the wire ($i, j$),  including subelements therein, is exposed to an initial uniform magnetic field $H_{ext,wire}^{k+1}(i, j)$ that is generated from the transport current circulating in the other wires in addition to the background magnetic field, i.e., $H_{ext,sub}^{k+1,1}(i, j, l)=H_{ext,wire}^{k+1}(i, j)$ with $l=1 \ldots N_{sub}$. Then, the component of the current density associated to the magnetic field $\textbf{J}_{H}^{k+1,1}$ is calculated subelement after subelement (one at a time). In addition, the component of current density associated to the transport current $\textbf{J}_{I}^{k+1,1}$ distributed in the entire region of a wire with coupling subelements is calculated by the electromagnetic $\textbf{A}-V$ formulation. It is worth noting that both $\textbf{J}_{H}^{k+1,1}$ and $\textbf{J}_{I}^{k+1,1}$ are calculated on the basis of resistivity $\bm{\rho}^{k}$ as a function of total current density $\textbf{J}^{k}$ at $k$ time step. The total current density $\textbf{J}^{k+1,1}=\textbf{J}_{H}^{k+1,1}+\textbf{J}_{I}^{k+1,1}$, resistivity $\bm{\rho}^{k+1,1}$ at all grid points and the net current in each subelement $I_{net}(i, j ,l)$ are then updated. After the first iteration ($m=1$), the external magnetic field $H_{ext,sub}^{k+1,2}(i, j, l)$ at each subelement is updated by the net currents of subelements obtained at the first iteration. The second iteration ($m=2$) is performed following a similar procedure as for $m=1$. Such iteration procedure for updating $H_{ext,sub}^{k+1,m}(i, j, l)$, $\textbf{J}_{H}^{k+1,m}$, $\textbf{J}_{I}^{k+1,m}$, $\textbf{J}^{k+1,m}$ and $\bm{\rho}^{k+1,m}$ is stopped once the maximum error between the external magnetic field of subelements $err$ or the error of the Joule heating is sufficiently small. We use an error threshold of 0.1\% for the Joule heating and 2.5 \% for the magnetic field of subelements. Eventually, the current density $\textbf{J}^{k+1}$, resistivity $\bm{\rho}^{k+1}$ magnetic field $\textbf{H}^{k+1}$, and Joule heating distribution over the entire cross-section of the coil is obtained for the time step $k+1$. As shown in Fig. S10 of the Supplementary Information, the convergence of the iteration depends on the number of turns of the coil. Two iterations are sufficiently accurate for small coils (less than 10$\times$10 turns) and one iteration is good enough for large coils.  In order to avoid divergences induced by the strong nonlinear $E-J$ constitutive relation, the Runge-Kutta method with variable time step is implemented to solve the electromagnetic equations.

\begin{figure*}
\centering
\includegraphics*[width=0.8\linewidth,angle=0]{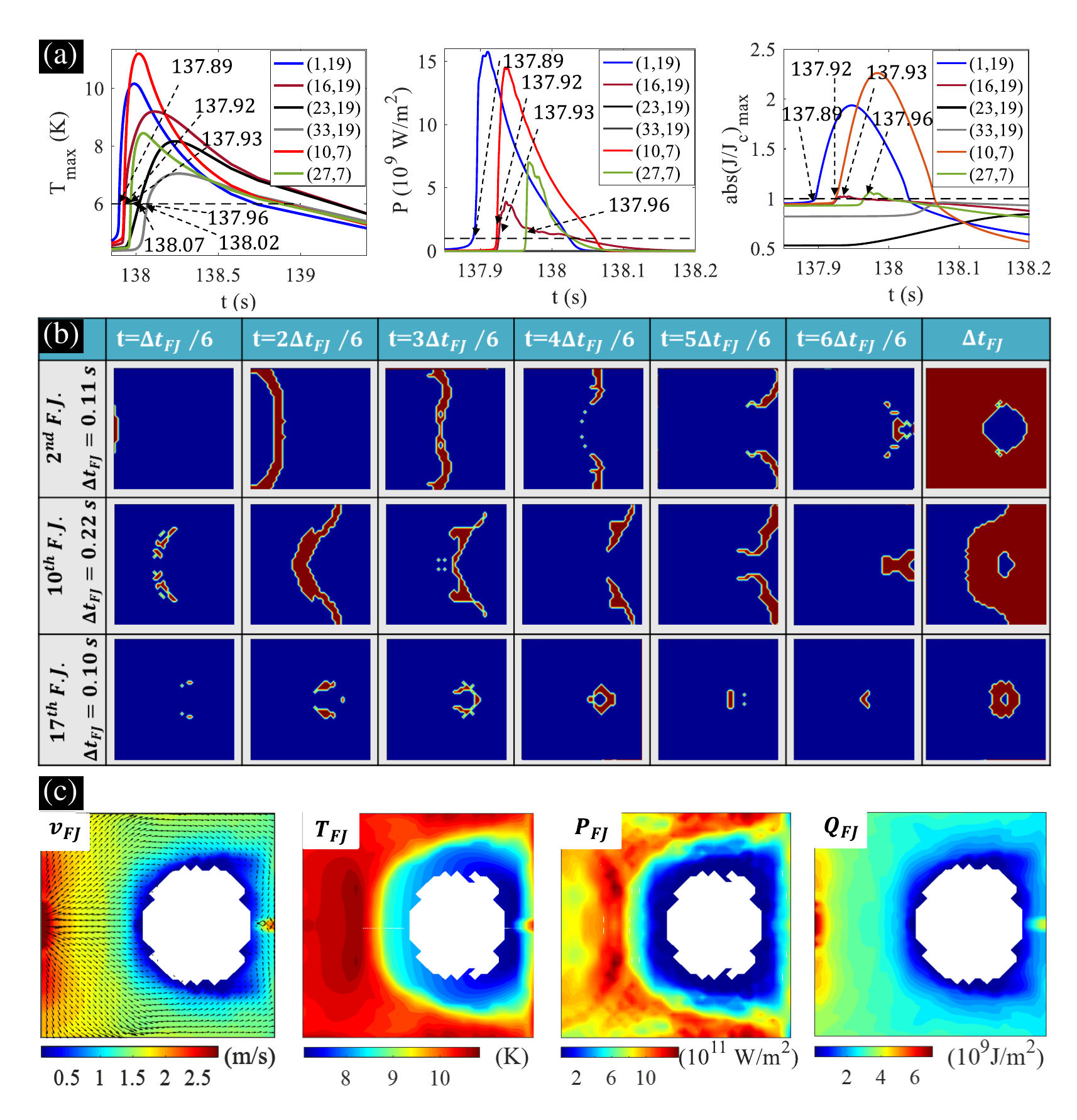}
\caption{(a) Time-evolutions of maximum temperature $T_{max}$, Joule heating power density $P$, and maximum normalized critical current density $J$ of wires at different partial locations during the second flux jump. (b) Snapshots of the flux jumping regions at six different moments (columns) and for three different flux jumps (rows). The panels on the last column show the propagation regions of the flux jumps. (c) Velocity field during the second flux jump where the arrows indicate the propagation direction. The right three panels show temperature $T_{FJ}$ and power density $P_{FJ}$ of each wire at the onset of the second flux jump, along with the quantity of Joule heating $Q_{FJ}$ generated within the time ranging from occurrence to peak for second flux jump.}
\label{Fig:4}
\end{figure*}

\begin{figure}
\centering
\includegraphics*[width=1.0\linewidth,angle=0]{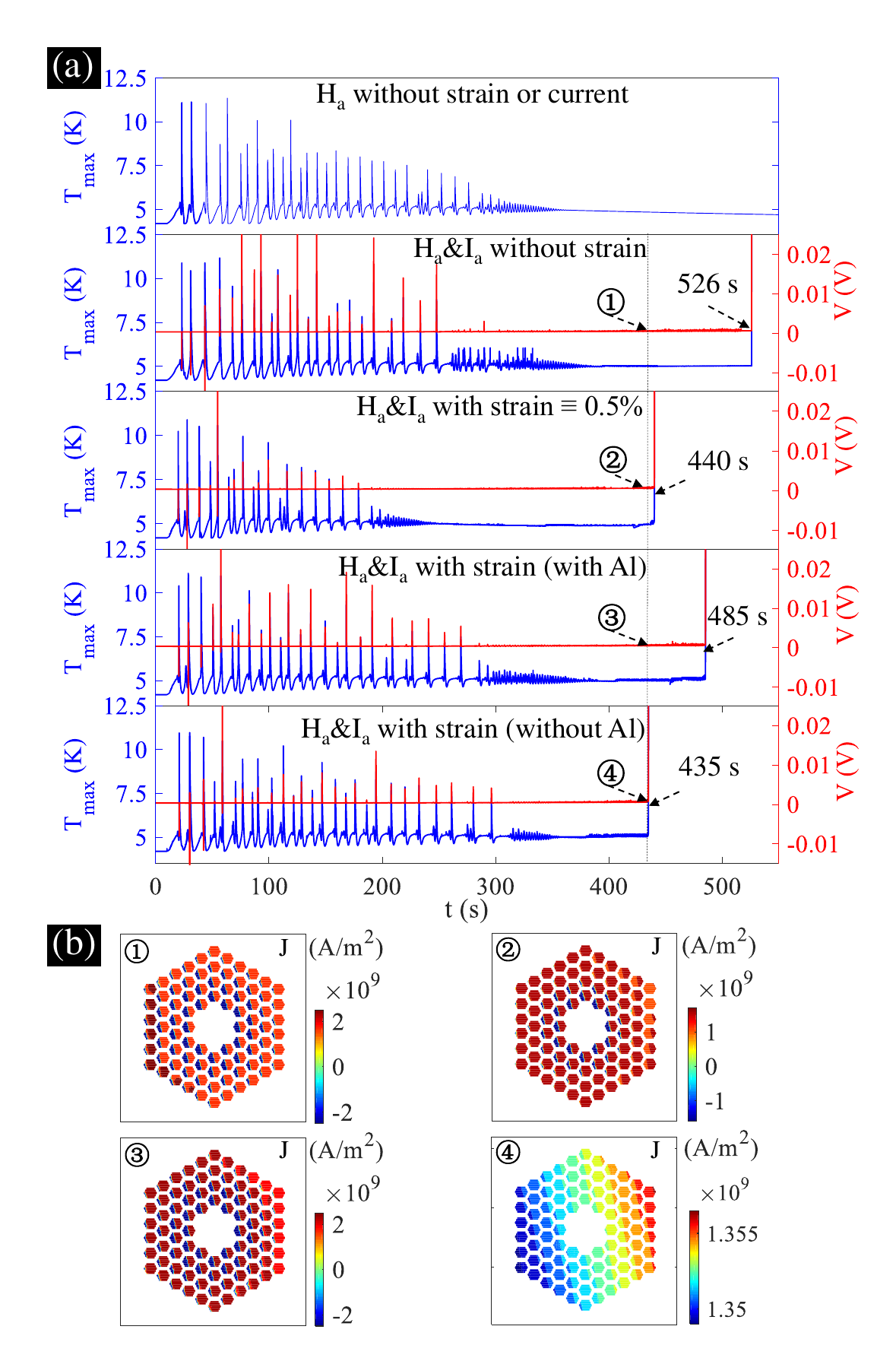}
\caption{(a) Simulated variations of the maximum temperature $T_{max}$ (blue) and terminal voltage (red) of a coil with $20 \times 20$ turns of Nb$_3$Sn wire for five different cases indicated in each panel. (b) The current density distribution in one of the wires of the coil for cases 2-5 at the time indicated by dashed line in panel (a).}
\label{Fig:5}
\end{figure}

\begin{figure*}
\centering
\includegraphics*[width=0.8\linewidth,angle=0]{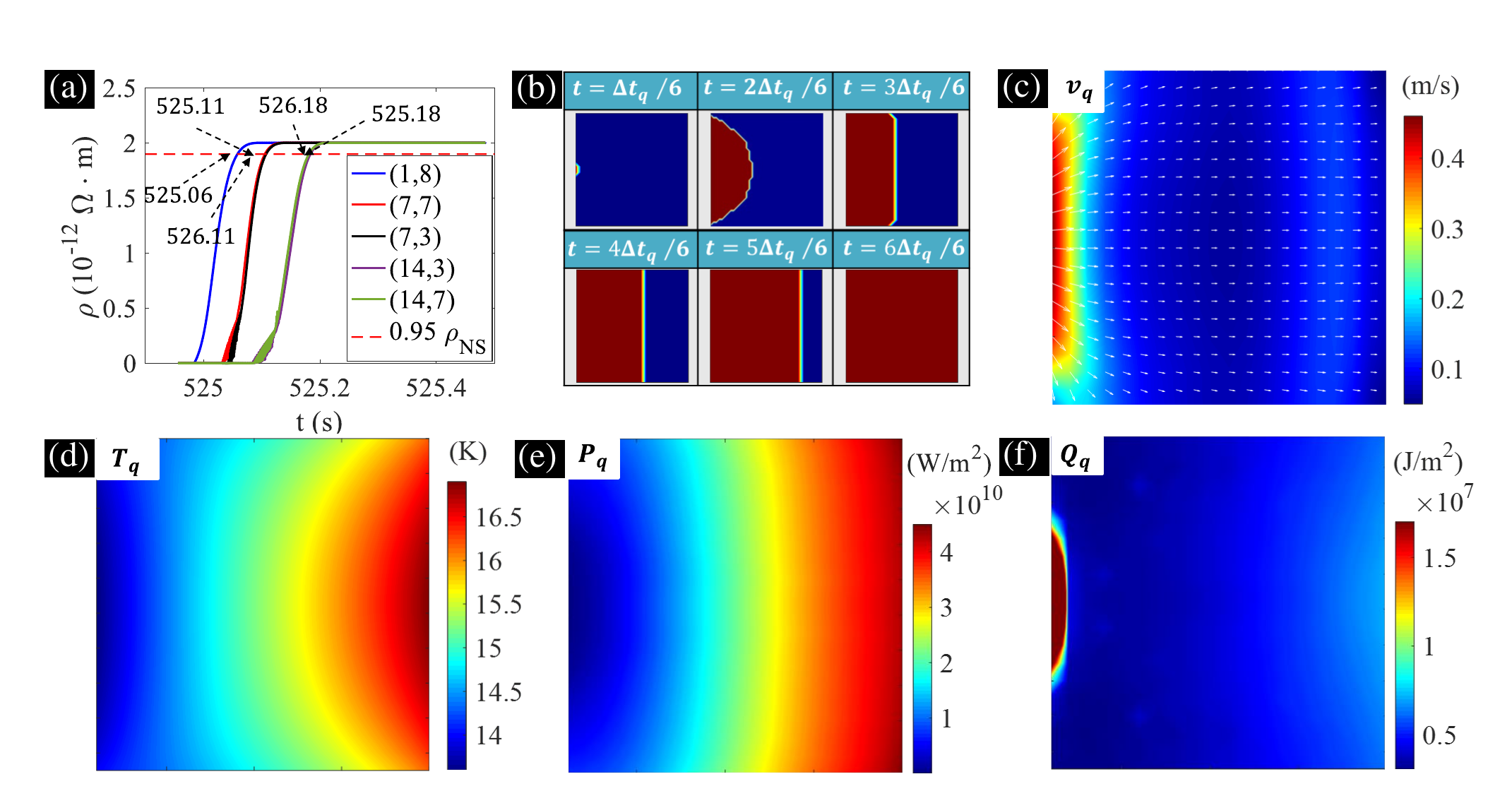}
\caption{(a) Resistivity as a function of time during a flux jump at different locations. The criterion of 0.95$\rho_n$ used to determine the quenching time for each wire of the coil is indicated with dashed red line. (b) Time-evolution of quenched regions (red colored) at six instances for case 2 as described in Fig. 5. (c) Velocity field in the coil during the quench with arrows indicating the propagation directions of the quench. (d-f) The instantaneous temperature $T_q$ and Joule heating power density $P_{q}$ of each wire at quench time, and the quantity of Joule heating $Q_{q}$ of each wire generated from the onset of rapid increase of $\rho$ until the quench.}
\label{Fig:6}
\end{figure*}

The temperature in the coil at each time step is obtained by the heat diffusion equation $c \frac{\partial T}{\partial t} = \bigtriangledown \cdot (\kappa \bigtriangledown T) + \boldsymbol{E} \cdot \boldsymbol{J}$ where $\boldsymbol{E} \cdot \boldsymbol{J}$ is the Joule heating source. This equation can be solved by considering the heat exchange boundary conditions at four borders on the cross-section of the coil, $-\kappa\left ( \bigtriangledown T\cdot  \mathbf{n} \right )  = h\left ( T-T_{0} \right )$, where $c$, $\kappa$, $h$ are the specific heat, thermal conductivity and heat transfer coefficient, respectively. The thermal parameters are assumed to be proportional to $T^3$, i.e., $c=c_{0} \left ( T/T_{0}  \right ) ^3$, $\kappa=\kappa_{0} \left (  T/T_{0} \right ) ^3$, $h=h_{0} \left ( T/T_{0}  \right ) ^3$. The alternating direction implicit (ADI) method is used to solve the heat diffusion equation in the composite coil consisting of Nb$_3$Sn, copper and epoxy. The above numerical algorithm for the coupled electromagnetic equations and heat diffusion equation is realized by a home-made code on C and CUDA programming language, which is executed on GPUs (GeForce RTX 4090 and 3090). Details concerning the flow chart of the main program, the electromagnetic $\textbf{A}-V$ formulation, the error as a function of iterations, the validation of separate method of calculating current density for a current-carrying wire exposed to $H_a$, and the parallel processing of the numerical algorithm on GPU, are presented in section IV of the Supplementary Information.

The constitutive relation between current and electric field for superconductors, $E=\rho J$, needs to invoke a nonlinear $\rho(J)$ which is mainly determined by the magnetic flux dynamics. In the past decades, various models describing the flux dynamics have been proposed, such as the Bean critical state model \cite{Bean}, the Anderson-Kim flux creep model \cite{flux-dynamics1,flux-dynamics2}, and the flux-flow model \cite{flux-dynamics3}. The flux dynamics in regimes spanning from the superconducting state to the normal state remains a subject of intensive study due to its sensitivity to temperature, strain, current, pinning nature, and magnetic field. A detailed discussion concerning the $E-J$ models is beyond the scope of the present work. Here, we adopt a $E-J$ law able to properly describe the electromagnetic response of superconductors including the flux creep (FC) state, the flux flow (FF) state and eventually the normal (N) state. In general, the critical current density $J_c$ (a parameter entering in the relation $\rho(J)$) also depends strongly on temperature $T$, strain $\varepsilon$, and magnetic field $H$. Fig. 2(d) shows the variation of $J_c$ with magnetic field obtained experimentally. A fitting curve based on the Kramer scaling law \cite{E-J} is used in the numerical simulations. The creep exponent $n$ (another parameter in $\rho(J)$) also varies with $T$ and $H$. The complex $E-J$ dependence with $\rho$, $J_c$, and $n$ for IT Nb$_3$Sn, and the thermal parameters for Nb$_3$Sn, cooper and epoxy are described in section III of the Supplementary Information. 

The parameters used in the numerical simulations were obtained from experimental measurements. To that end, we prepared a short sample of IT Nb$_3$Sn wire, same as the one used in the solenoid coil, with both ends polished (see Fig. 2(c)). The sample is exposed to a cycling transversal magnetic field with sweeping rate of 0.01 T/s at 4.2 K. Figure 2(d) shows that the simulated magnetization nicely captures the features observed in the experimental loop. This indicates that the chosen electromagnetic and thermal parameters are suitable for the wire used in the solenoid and validates the proposed numerical algorithm. Both experiments and simulations show that the magnetization of the Nb$_3$Sn wire does not decrease to zero during the flux jumps, suggesting that the temperature does not exceed $T_c$ during partial flux jumps. The simulations reveal that the maximum temperature achieved during the flux jump is about 11 K (see Fig. S6 in Supplementary Information) whereas the current density decreases significantly during this process (see Fig. 2(d)).

\large
\vspace{4ex}
\noindent
\textbf{{Results and Discussions}}

\normalsize
\noindent
\textbf{Flux Jumps Propagation in a Coil}

Encouraged by the success of the proposed numerical algorithm, we then explored the flux jumps in a solenoid coil with 1600 (40 $\times$ 40) turns of Nb$_3$Sn wires. The upper panel of Fig. 3(a) shows the experimentally observed voltage signal exhibiting frequent flux jumps during a continuously current ramp of 0.5 A/s for the solenoid coil. Due to the fact that $J_c$ of Nb$_3$Sn is very sensitive to strain, this effect should also be taken into consideration in the numerical simulations. The mechanical response of the solenoid coil includes three parts: thermal strain caused by cooling down to 4.2 K, pre-strain process caused by the aluminum strip and the electromagnetic strain produced by the Lorentz force. Detailed analyses of the mechanical deformation and $J_c(\varepsilon)$ are shown in Section III and Section V of the Supplementary Information. The lower two panels of Fig. 3(a) indicate that the first flux jump is triggered after 90 s, subsequently the solenoid undergoes a train of jumps until~ 1100s after which it remains stable without flux jumps. This result agrees well with the one obtained from the experimental test. The difference of number of flux jumps between simulations and experiments may result from the departure of ideal defect-free Nb$_3$Sn wire assumed in the numerical model.

Figure 3(b) shows the number of flux jumps across the entire coil during the ramping process. The statistics of the flux jumps in each wire during this process reveals that the flux jumps are not triggered uniformly in all wires. The thermomagnetic instabilities are statistically less likely to occur in the center of the right region. This is because the ramping rate of the local magnetic field in this region is substantially smaller than elsewhere. Fig. 3(c) shows  snapshots of the current density, the magnetic field distributions in subelements, and the temperature distribution in the coil during the second flux jump. The lower panels shows that full flux penetration is achieved in the outer wires while the inner wires are only partially penetrated by the magnetic flux. Furthermore, the temperature is nearly uniform in each wire whereas a large temperature gradient can be observed at the interface of each wire.

The most fascinating aspect of the phenomenon under consideration, concerns the nucleation process of flux jumps and the subsequent growth and propagation throughout the coil. In order to address this question, a criterion is needed to discern whether a thermomagnetic instability has been triggered in one particular wire. As shown in Fig. 4(a), the temperature rises in all of six wires chosen at different locations. However, the time-evolution of $T$ does not represent a reliable criterion because the heat conduction from surrounding wires can also lead to a local increase of temperature. As an illustration of this point, in the middle panel of Fig. 4(a) the Joule heating power density for two wires [(23, 19) and (33, 19)] is plotted. One remarks that the dissipated power is very small during the flux jump process, thus indicating that the flux jump does not occur in these two wires, even though the temperature has increased. Alternatively, the rightmost panel shows that $J$ is always less than $J_c$ in those wires without flux jumps. Based on these considerations, we adopt the criterion $|J/J_c|>$1 as the threshold indicating the nucleation of a flux jump.

Fig. 4(a) further indicates that flux jumps do not occur in different wires at the same time. Fig. 4(b) shows the flux jumping regions (red colored) at six different moments and for three different flux jumps taking place chronologically. One can see that in an early stage (upper row), the flux jumps are triggered on the left side (corresponding to inner radius of the coil), while in a later stage, the flux jumps are firstly observed in inner wires (central and lower row). Interestingly, for the latter, the flux jumps do not propagate towards the outer rim of the coil, instead the region of flux jumps remain spatially confined because $J_c$ is weakened by the high magnetic field in the outer region. The left panel of Fig. 4(c) shows that the propagation velocity field of the second flux jump is nonuniform over the coil and lies within a range of 0.1-2.8 m/s in agreement with previous experimental measurements \cite{FJ-speed}. In order to explore what determines the propagation velocity distribution in the coil, we calculated the maximum temperature $T_{max}$, the maximum Joule heating power density $P_{max}$ of each wire during the second flux jump, and the quantity of Joule heating within the time-range from occurrence to peak of second flux jump $Q_{FJ}$ in each wire. It is surprising that the propagation velocity of the flux jump from a wire to its neighbouring wire is mainly determined by the $Q_{FJ}$ rather than $T_{max}$ or $P_{max}$. Moreover, the propagation directions of the flux jump are mainly related to the gradient of $Q_{FJ}$, which indicates that the flux jump of a wire preferably propagates to its neighbouring wire with larger $Q_{FJ}$. As a consequence,the flux jump cease from propagation to the wire has no sufficient potential energy released. Animations illustrating the propagation of 2nd, 10nd and 17th flux jumps can be seen in the Supplementary Movies 1-3.

\vspace{4ex}
\noindent
\textbf{Quenches Propagation in a Coil}

Let us now scale up the problem and explore the time-evolution of quenches in a coil with $20 \times 20$ turns. To that end, we consider five different cases, each with a progressive increase of complexity. The coil is exposed to a non-uniform self-field generated by a transport current of 2 A/s and a uniform background magnetic field of 0.015 T/s. In case 1, both transport current and strain effect are neglected in the numerical simulation. In case 2, the transport current is taken into consideration, whereas a constant strain $\varepsilon=0.5\%$ for each wire, real thermal and electromagnetic strain fields with and without pre-strain are considered in cases 3-5, respectively. As shown in Fig. 5(a), the voltage spikes in cases 2-5 become progressively smaller after the middle stage of the entire current loading phase, and eventually vanish with increasing the transport current, which is consistent with the experimental results shown in Fig. 3(a). Comparing with cases 1 and 2, one can see that the transport current with low ramping rate has almost no impact on the threshold value of virgin flux jump and the frequency of flux jumps. Cases 2-5 indicate that strain causes a significant premature quench, likely because strain leads to a serious degradation of $J_c$. Therefore, taking into consideration strain effects is a critical issue for coil design. Indeed, comparing cases 3 and 4, suitable pre-strain by the aluminum strip can significantly improve the quench current. As shown in Fig. 5(b), the current density in all subelements exhibits full current-like state and almost reaches up $J_c$ at the moment when quench occurs in case 5, while current density in some subelements is still in field-like state and thus these subelements still have capacity for more transport current.

The next challenge consists in identifying a reliable indicator for the quench propagation in the coil. As shown in Fig. 6(a), the resistivity of each wire increases rapidly to its normal state value $\rho_{n}$. Thus, we choose $\rho > 0.95 \rho_n$ as the quench criterion for each wire. Fig. 6(b) shows that the onset of quench appears at the center of the left border and it propagates towards the right border until all wires of the coil switch to the normal state. From Fig. 6(c), one can see that the velocity of quench propagation is not uniform in the coil and the quench propagate much more rapidly in the left region than elsewhere. Comparing the velocity field of quench propagation with the time-integration of Joule heating $Q_q$ from the onset of rapid increase of $\rho$ up to the quench (see Fig. 6(d)), instantaneous Joule heating power density $P_q$ (Fig. 6(e)) and instantaneous temperature $T_q$ at quench time, we demonstrate that the propagation velocity of the quenching process is undoubtedly related to $Q_q$ of each wire. The dynamic propagation of a quench can be found in Supplementary Movie 4.

\large
\vspace{4ex}
\noindent
\textbf{{Summary}}
\normalsize

In summary, we have developed a parallel numerical algorithm executed on GPU and permitting to deal with the correlated system of full-sized solenoid coil with thousand turns of multifilamentary superconducting wires. We have carried out experimental tests on a short sample of IT Nb$_3$Sn wire as well as on a solenoid coil. The simulated results reproduce the experimental data to a large extent. Moreover, utilizing the developed GPU algorithm, we were able to unveil the real-time dynamic and reveal detailed propagating velocity fields of magnetic flux jumps and quenches in superconducting coils. The most striking finding is that the velocity field of the thermomagnetic instability front is mainly related to the quantity of Joule heating rather than the Joule heating power or the maximum temperature. Although at present the numerical algorithm is intended for solenoid magnets, it can be easily extended to other structured magnet, such as racetrack coils. The large-scale GPU-advanced algorithm lays the foundation for the next-generation of numerical superconducting magnet techniques and provides a powerful tool for optimal design of future high-field magnets.

\large
\vspace{4ex}
\textbf{Data Availability}
\normalsize

The data that support the findings of this study are available from the corresponding author upon reasonable request.

\large
\vspace{6ex}
\noindent
\textbf{Acknowledgement}
\normalsize

\vspace{1ex}
We acknowledge support by the National Natural Science Foundation of China (Grant Nos. 12372210 and 11972298, 11427904). We thanks for the helpful discussions with Prof. You-He Zhou at Lanzhou University, and Dr. Peng Ma in Western Superconducting Technologies Co., Ltd.

\large
\vspace{6ex}
\noindent
\textbf{Author contributions}
\normalsize

C.X. designed the research, formulated the idea of solution, and conceived the main numerical algorithm. H.X.R and P.J. implemented the numerical simulations, algorithm validation schemes, analyzed the results and Q.Y.W. performed the mechanical calculations under the supervision of C.X. W.L. prepared the short samples and P.J. implemented experimental measurements for short wires. L.T.S. X.J.O. and W.L. fabricated the solenoid coil and implemented measurements. C.X., H.X.R. prepared the first draft of the manuscript with contributions from A.V.S. All authors contributed to discussions and revision of the manuscript to its final version.

\large
\vspace{6ex}
\noindent
\textbf{Competing interests}
\normalsize

The author declares no competing interests.

\end{document}